\documentclass[aps,prl,reprint,groupedaddress,amsmath,amssymb, floatfix]{revtex4-1}
\usepackage{tikz}

\usepackage{graphicx}
\usepackage{dcolumn}
\usepackage{bm}
\usepackage{comment}
\usepackage[normalem]{ulem}

\usepackage{tabularx, booktabs}
\newcolumntype{C}{>{\centering\arraybackslash}m{3.5em}}
\newcommand{\av}[1]{\langle #1 \rangle}

\newcommand{\del}[1]{}
\newcommand{\new}[1]{#1}

\usepackage{tikz}

\begin{document}

\preprint{APS/123-QED}

\title{Generalization of Turbulent Pair Dispersion to Large Initial Separations}
\author{Ron Shnapp}
\email{ronshnapp@mail.tau.ac.il}
\author{Alex Liberzon}%
 \email{alexlib@eng.tau.ac.il}
\affiliation{%
 School of Mechanical Engineering\\
 Tel Aviv University, Tel Aviv 69978
}%
\collaboration{International Collaboration for Turbulence Research} 

\date{\today}

\begin{abstract}

We present a generalization of turbulent pair dispersion to large initial separations ($\eta < r_0 < L$), by introducing a new time scale, $\tau_{v_0}$, that reflects the persistence of initial conditions at time $\tau=0$. Results of 3D Lagrangian tracking experiments at moderate Reynolds numbers show that pairs, for which the new time scale is shorter than the eddy turnover time scale, separate as in the Richardson superdiffusive regime, $\av{\Delta r^2} \propto \tau^3$. The analysis of delay times \new{(time interval to cross $\Delta r = \rho \, r_0$)} of these conditionally sampled pairs exhibit $\rho^{2/5}$ scaling. 

\end{abstract}

\pacs{Valid PACS appear here}
\maketitle

\new{Consider two fluid particles in a turbulent flow that at some moment are at a close proximity to each other. As time progresses, the separation between them would grow. The statistics of the change in separation distance between such fluid particles is termed pair dispersion. Pair dispersion and the two-point description of fluid velocities underline central theories of turbulence. This is because of the spatial and temporal correlations of two points that are intrinsic in turbulent flows. These correlations differ turbulence from the Brownian motion case and make it scale dependent, with different scaling exponents for the moments of separation, as the distance between the pairs grow.}

\new{We denote the separation distance between particles as $r(\tau) = |\boldsymbol{x}_1(\tau) - \boldsymbol{x}_2(\tau) |$ ($\boldsymbol{x_{1,2}}$ are positions of particles 1,2, $\tau = t - t_0$, $t_0$ is an arbitrary time instant, $| \cdot |$ is the $L_2$ norm, and $r_0 \equiv r(t_0)$ ).} Richardson~\cite{Richardson1926} studied pair dispersion from a small source ($r_0 \rightarrow 0$) semi-analytically, and found that the ensemble averaged $\av{r^2}$ is superdiffusive, i.e. $\av{r^2} \propto \tau^3$. Later, Batchelor~\cite{Batchelor1952} suggested that pairs separate first ``ballistically'', up to the eddy turnover time at a scale $r_0$, $\tau_0 = \left(r_0^2 / \epsilon\right)^{1/3}$ ($\epsilon$ is the mean rate of turbulent kinetic energy dissipation). These solutions can be presented in the form of $\Delta r = (r(\tau)-r_0)$ as recently reviewed by~\cite{Salazar2008}:

\begin{equation}
\langle \Delta r^2 \rangle = \begin{cases}
\frac{11}{3}C_2 (\epsilon r_0)^{2/3} \tau^2  \quad \quad  &   \text{for} \,\, \tau \ll \tau_0 \\
g \epsilon \tau^3  &  \text{for} \,\,  \tau_0 \ll \tau \ll T_L
\end{cases}
\label{eq:separation_laws}
\end{equation}
\noindent where $C_2$ is the Kolmogorov constant (see definition below), $T_L$ is the Lagrangian integral time scale, and $g$ is the Richardson constant, which was obtained by $g \approx 0.5 \pm 0.05$~\cite{Julien1999,Ott2000,Boffetta2002,Biferale2005}.
Numerical simulations and experiments~\cite{Vanderwel2014,Ni2013,Boffetta2002,Julien1999} provided supporting evidence for the laws in Eq.~\eqref{eq:separation_laws} for small and finite initial separations $r_0 \sim \eta$. \new{Ref.~\cite{Biferale2005} have shown that pairs with $r_0<2.5\eta$ in a high Reynolds number direct numerical simulations (DNS), separate superdiffusively for some time interval, however for larger $r_0$, this regime disappears.} Using DNS, Refs.~\cite{Biferale2014,Scatamacchia2012} have shown the complexity of the process that mixes statistics of extreme events - from very fast spreading pairs to very slow separation rates.

\del{Pair dispersion has also been studied}For the case of large initial separations -- $r_0 \gg \eta$, a consensus regarding the validity of Eq.~\eqref{eq:separation_laws} is lacking. Refs.~\cite{Ott2000,Berg2006} identified a superdiffusive regime in particle tracking experiments by introducing a ``virtual time origin'' concept. Refs.~\cite{Bourgoin2006,Ouellette2006}, based on high Reynolds number experiments, show that the ballistic regime dominates the separation scaling for the entire range of times and scales studied. On one hand, Ref.~\cite{Biferale2005} using ``exit times'' analysis (statistics of time to reach a set of thresholds $r_{n+1} = \rho r_{n}$, with $\rho > 1$~\cite{Boffetta2001}) found supporting evidence for the superdiffusive regime in a narrow range of scales. On the other hand, Ref.~\cite{Ouellette2006} have shown using their experimental data that the ``exit time'' analysis does not support \del{the laws in Eq.~\eqref{eq:separation_laws}}\new{a superdiffusive regime of pair separation}.



%
\begin{figure*}[t]
	\begin{tikzpicture}
	\node[inner sep=0pt] (a) at (0,0)    
	{\includegraphics[width=0.5\textwidth]{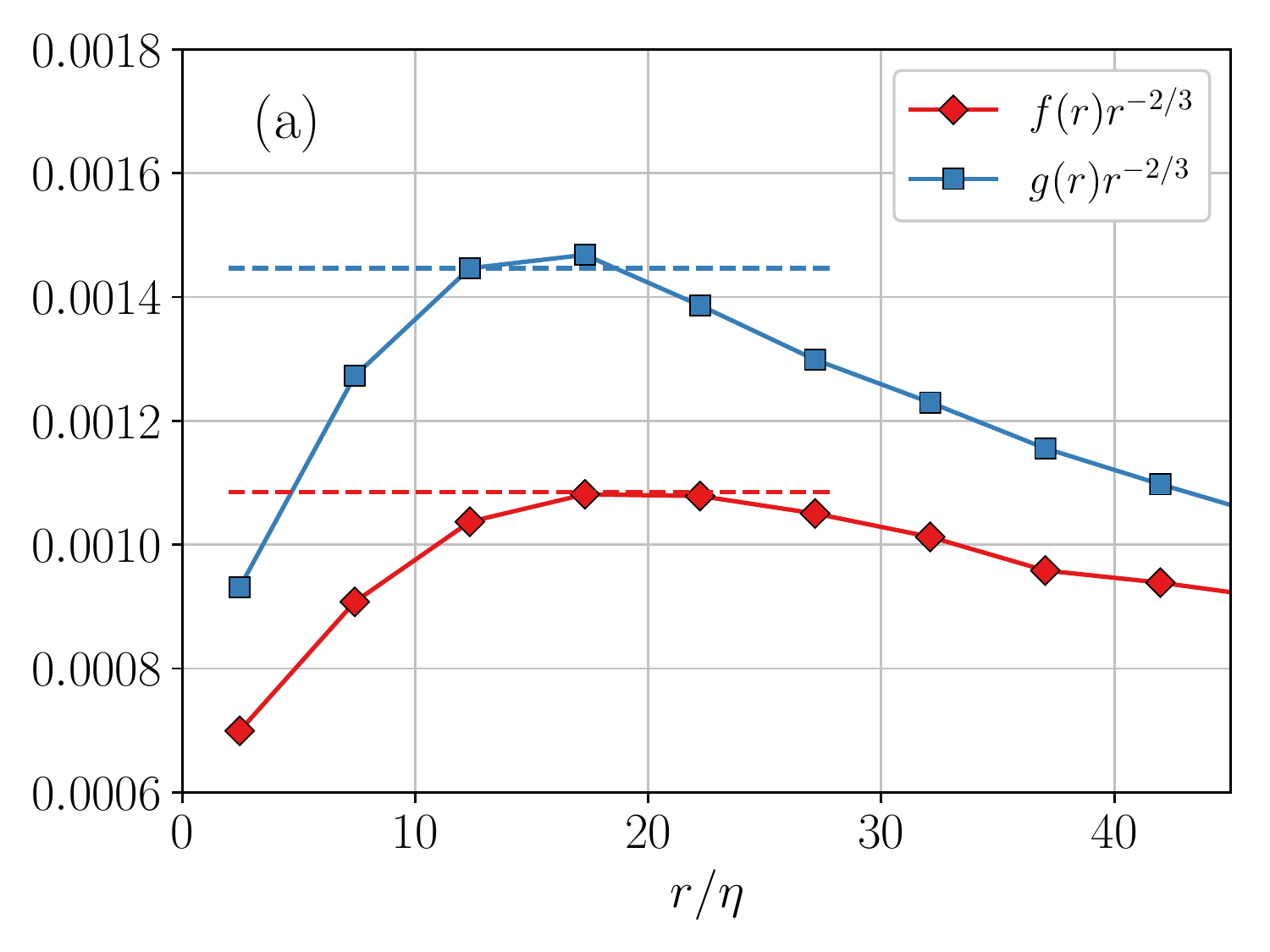}};
	\node[inner sep=0pt] (b) at (8.85,0) 
	{\includegraphics[width=0.5\textwidth]{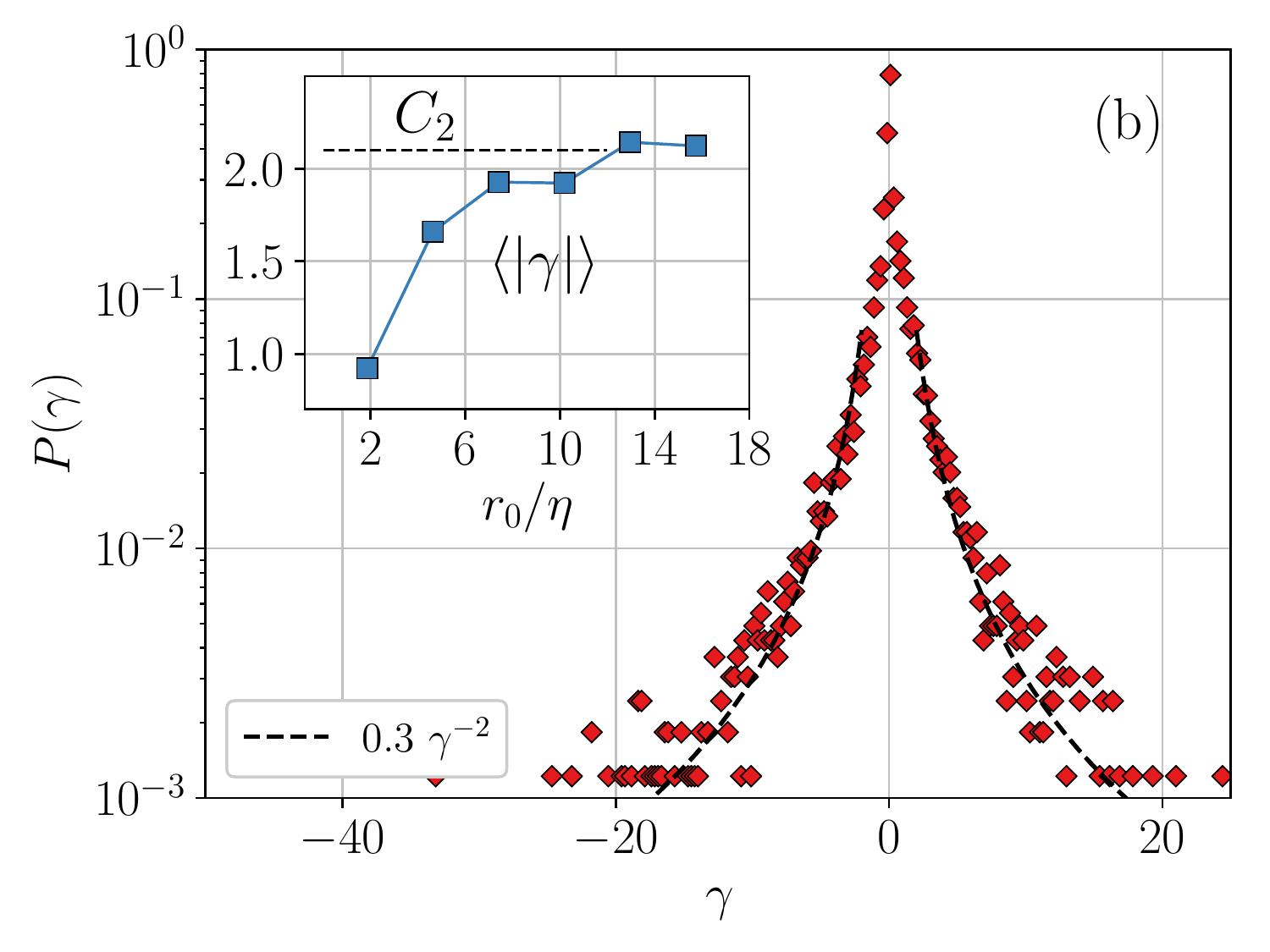}};
	\end{tikzpicture}

	\caption{ (a) Eulerian second order structure function in the longitudinal ($f(r)$) and transverse ($g(r)$) directions, compensated by $r^{2/3}$ according to inertial subrange scaling. (b) Main: distribution of the time scales ratio $\gamma$ (Eq.~\eqref{eq:Time_scale_ratio}) for all the pairs of particles that are initially separated by distance $r_0<10$ mm. Dashed lines are proportional to $\gamma^{-2}$. Inset: ensemble average of absolute value of $\gamma$ as a function of $r_0/\eta$. \label{fig:vel_hist}}
\end{figure*}

In this Letter we suggest a generalization of pair dispersion laws for the case of large initial separations, $\eta \leq r_0 \leq L$ ($L$ is an integral length scale). We introduce a new dimensionless parameter $\gamma=\tau_{v_0}/\tau_0$ defined as the ratio between the time scale of separation ($\tau_{v_0}  = {v_{||,0}^2}/{\epsilon}$, $v_{||} = dr/dt$ is a pair separation velocity, $v_{||,0} = v_{||}(t_0)$), and the eddy turn-over time $\tau_0$, at the scale $r_0$. Using results from a three-dimensional particle tracking (3D-PTV) experiment of pairs in the range of $2\eta \leq r_0 \leq 20\eta$, we demonstrate that pairs with small $\gamma$ separate superdiffusively, whereas pairs with high $\gamma$ separate ballistically:
\begin{equation}
\langle \Delta r^2 \rangle \propto \begin{cases}
\tau^2  \quad \quad  &   \text{for} \,\, |\gamma| \gg 1 \\
\tau^3  &  \text{for} \,\,  |\gamma| \ll 1
\end{cases}
\label{eq:gamma_sep}
\end{equation}
\new{This experimental observation is consistent with DNS results at $Re_\lambda\approx 400$ (see supplementary material).} We argue that \new{the observed Eq.~\eqref{eq:gamma_sep} serves as a generalization for the the supperdiffusive separation case of pairs with small initial separations~\citep{Biferale2005}.}\del{is consistent with our description.} This is because of an increased probability for small $\gamma$ values to occur with $r_0 \to \eta$, as we show experimentally \new{and confirm using DNS (see supplementary material).} We observe that Lagrangian correlations of pair relative velocity is stronger with $\gamma$ increasing from negative to positive. For all the pairs we detect an asymptotic tendency of mean ``delay times''~\cite{Rast2011} towards a $\rho^{2/3}$ scaling law, while for the pairs with $|\gamma| < 0.1$ we observe a scaling law with exponent $2/5$.

\medskip
We use a classical experiment of two oscillating grids in a water tank creating zero mean shear turbulence~\cite{Ott2000, Zeff2003}. Two rigidly connected grids (105 mm vertical separation) are oscillated vertically with an amplitude of 10 mm at a rate of 7 Hz ($\approx 2.4 \times \tau_\eta^{-1}$)\new{, within the water tank of cross section $300\times 300$ mm$^2$.} We apply 3D-PTV~\cite{Dracos:1996} to measure the turbulent flow using four high speed cameras and an unique real-time image processing on-hardware system~\cite{Kreizer:2011}, combined with a dedicated open source software~\cite{openptv,Meller2016}. We obtained a large dataset of particle trajectories of polyamide spheres ($d_p \approx 50\, \mu m$), in a region $80 \times 60 \times 40$ mm$^3$ with position accuracy $\leq 300 \, \mu$m, from which velocities were calculated based on the method presented in Ref.~\cite{Luthi2005}. \del{The flow velocity field in the measurement volume is quasi-homogeneous and quasi-isotropic, with ensemble averaged root mean square of turbulent velocities of 12 mm/s, and a small residual secondary flow  of about 2 mm/s, which was verified not to affect the pair separation velocity statistics}\new{The flow velocity field in the measurement volume has ensemble averaged root mean square of turbulent velocities of $u'=$12 mm/s, and a small residual secondary flow  of about 2 mm/s. There is also a small degree of anisotropy in the direction of the grid oscillations. Yet, as is shown below, the turbulence structure is quasi-homogeneous and quasi-isotropic in the sence that the flow is consistent with Kolmogorov similarity predictions.}

%

Pair dispersion is closely related to the longitudinal $f(r)$ and transverse $g(r)$ velocity structure functions~\cite{Monin2007}, defined from the relative velocity of trajectory pairs, $\boldsymbol{v} \equiv \dot{\boldsymbol{x}}_2 - \dot{\boldsymbol{x}}_1 \equiv d\boldsymbol{r}/dt$~\cite{Pope2000, Ott2000} as:
\begin{eqnarray} \label{eq:calc_structure_func}
 f(r) & \equiv & \left \langle \, \left( \boldsymbol{v}(r) \cdot \boldsymbol{r}/r  \right)^2 \, \right\rangle \\
 g(r) & \equiv & \frac{1}{2} \left[\langle \boldsymbol{v}^2(r) \rangle - f(r) \right] \nonumber
\end{eqnarray}
%
We plot the structure functions compensated with $r^{2/3}$ in figure~\ref{fig:vel_hist}(a). The values at which both functions show a plateau (dashed horizontal lines) agree with the locally homogeneous value of $g(r) = 4/3 f(r)$~\cite{Pope2000}. \new{This is an indication that the flow is in good approximation locally isotropic turbulence, and that we should expect for Kolmogorov similarity arguments to be valid for our flow}. Using the relation $f(r) = C_2 (\epsilon r)^{2/3}$~\cite{Monin2007} and the value $C_2=2.1$~\cite{Sreenivasan1995}, the mean dissipation rate can be estimated as $\epsilon = 10$ mm$^2$s$^{-3}$. The corresponding Taylor-microscale Reynolds number $Re_\lambda \approx 180$, and the Kolmogorov length and time scales are $540 \, \mu$m and 0.34 s, respectively. \new{Furthermore, we can estimate the order of magnitude of the integral length scale as $L\sim u'^{3}/\epsilon =\mathcal{O}(100\,\text{mm})$}.
%
%

%
\begin{figure*}
\begin{tikzpicture}
\node[inner sep=0pt] (a) at (0,0)    
{\includegraphics[width=0.5\textwidth]{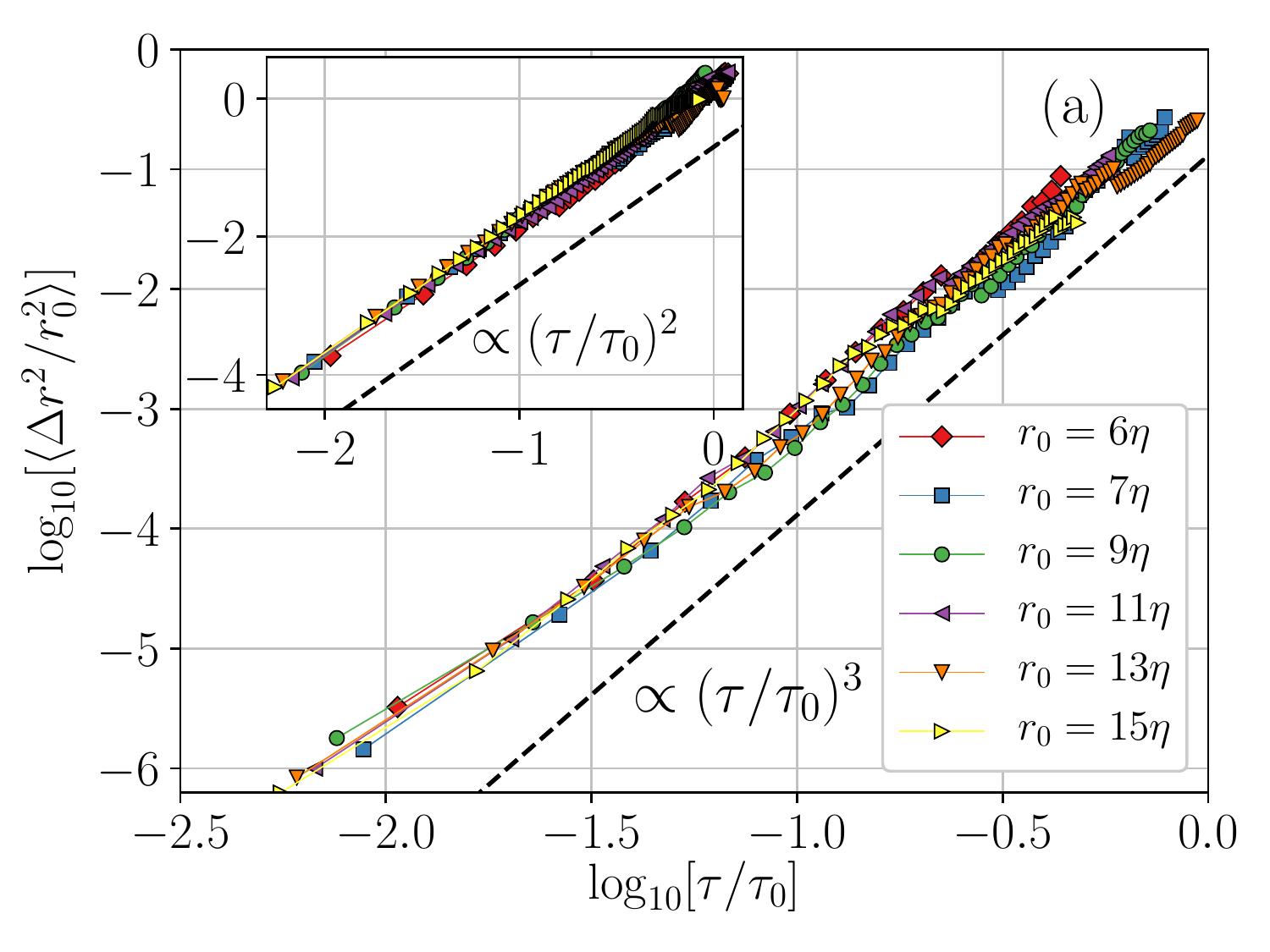}};

\node[inner sep=0pt] (b) at (8.85,0) 
{\includegraphics[width=0.5\textwidth]{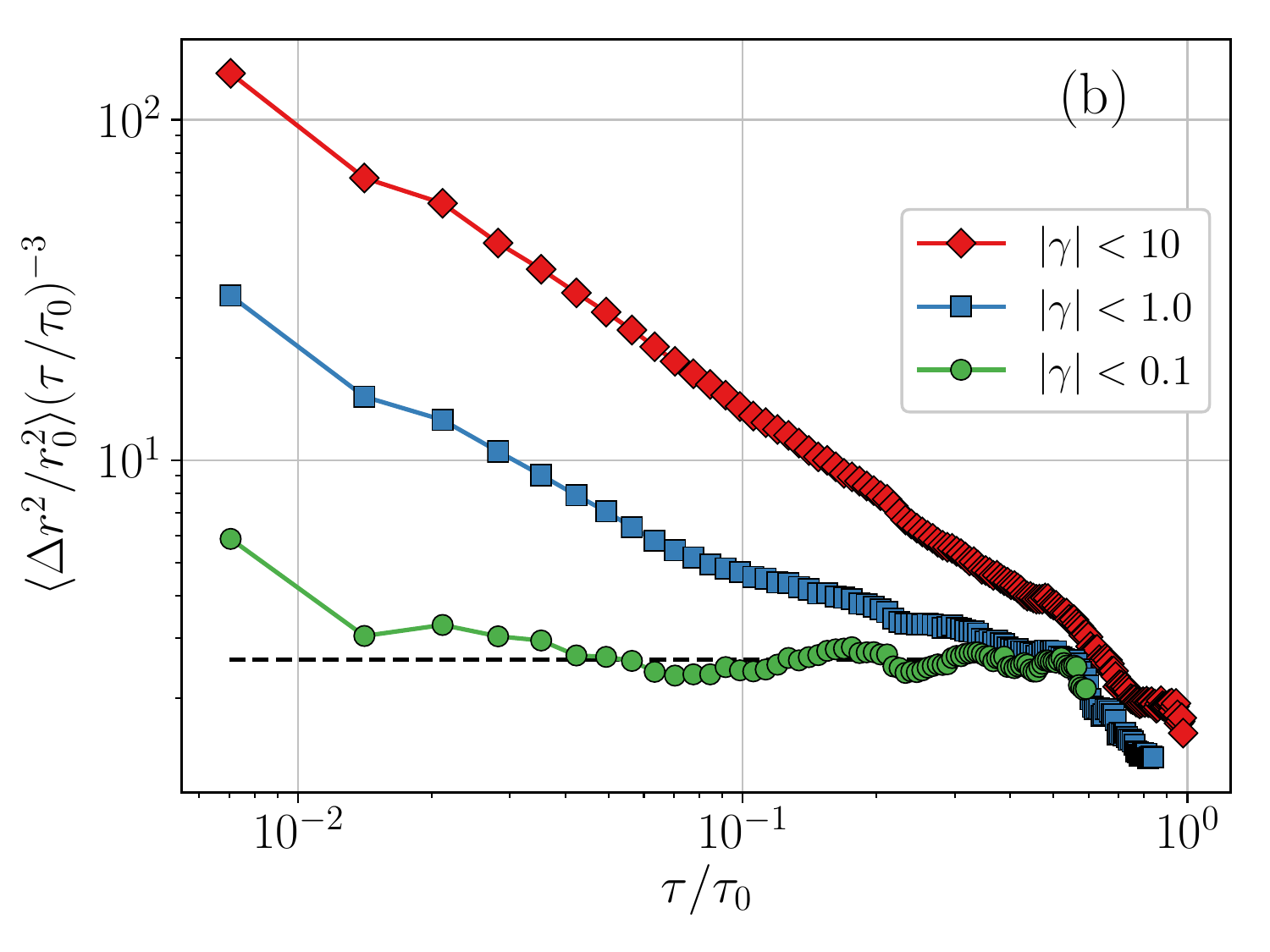}};
\end{tikzpicture}
\caption{(a) - Main: Second moment of the normalized change in separation distance for pairs of particles at various initial separation distances, conditioned for $|\gamma|<0.1$. The black dashed line shows a superdiffusive regime. Inset: same plot for unconditioned pairs with respect to $\gamma$. The black dashed line shows a ballistic scaling regime. (b) - Second moment of the change in separation distance compensated by the superdiffusive $(\tau/\tau_0)^3$ scaling, for pairs conditioned with $\gamma$ being smaller than three different threshold values.\label{fig:separation_curves}}
\end{figure*}

\medskip
We define a new time scale that relates to the initial conditions of pair separation, since pairs with $r_0 > \eta$ may experience a large verity of relative velocities. The new time scale should predict a typical time for change to occur in the \textit{initial} rate of separation, $v_{||,0}$. Kolmogorov scaling in the inertial range, $\tau_\eta \ll \tau \ll T_L$, predicts that the variance of Lagrangian velocity increments is proportional to $\epsilon \tau$~\cite{Monin2007}. Therefore we define the time scale as:
%

%
\begin{equation}
\tau_{v_0}  = \frac{v_{||,0}^2}{\epsilon} .
\label{eq:rel_vel_time_scale}
\end{equation} 
%
$\tau_{v_0}$ is the time scale during which a pair retains the order of magnitude of its initial relative velocity. It is in some sense analogous to the Batchelor~\cite{Batchelor1952} $\tau_0$, which predicts the change in the initially ballistic separation regime. The relation between $\tau_{v_0}$ and the separation process is emphasized through the dimensionless ratio, $\gamma$, defined in respect to $\tau_0$, 
\begin{equation}
\gamma = \frac{\tau_{v_0}}{\tau_0}  \mathrm{sgn}(v_{||,0}) = \frac{|v_{||,0}|\cdot v_{||,0}}{(r_0 \epsilon)^{2/3}}
\label{eq:Time_scale_ratio}
\end{equation}
The key parameter $\gamma$ defines the time to retain the initial rates of separation with respect to the rates of eddy breakdown at the scale $r_0$. Also, for $\gamma>0$, the particles are initially spreading, while for $\gamma<0$ they are initially getting closer. The probability density function (PDF) of $\gamma$ for the pairs at $\eta <r_0 < L$ is shown in figure~\ref{fig:vel_hist}(b). The distribution of $\gamma$ is symmetric, zero averaged and non-Gaussian with relatively high probability of $|\gamma| \gg 1$. 
In the inset of figure~\ref{fig:vel_hist}(b) the mean absolute value $ֿ\langle |\gamma|\rangle$ is plotted as a function of initial separation. Similarly to $f(r)$, the ensemble average $\av{|\gamma|} = \av{v_{||}^2}/(r_0 \epsilon)^{2/3} $ exhibits a plateau in  the range $\eta \leq r_0 \leq L$ and it is equal to $C_2$~\cite{Sreenivasan1995}. Therefore, as the initial separation of a pair, $r_0$, approaches towards $\eta$, lower values of $\gamma$ become more probable. These observations lead us to the notion that as $r_0$ becomes smaller, $|\gamma|$ tends to zero. \new{This was confirmed against high $Re_\lambda$ DNS results (see supplementary material).}

%

\medskip
Now we address the effect of $\gamma$ on the process of pair separation. We plot in figure~\ref{fig:separation_curves}(a) the time evolution of the average rate of squared change in separation $\av{\Delta r^2/r_0^2 }$ versus time normalized by $\tau_0$, for different initial separations. The main figure shows results for pairs of particles with $|\gamma|<0.1$, while in the inset we show curves for all the pairs. In both cases, the results for all inspected values of $r_0$ collapse on a single curve. The slope of the curves in the inset corresponds to a $(\tau/\tau_0)^2$ scaling of a ballistic regime. This is in agreement with previous works~\cite{Bourgoin2006,Ouellette2006,Ni2013}. In contrast to that, the statistics of pairs conditioned on $|\gamma|<0.1$ (neither spreading nor converging at the moment of selection), differ immensely.  After a very short time, almost two orders of magnitude less than $\tau_0$, $\av{\Delta r^2 /r_0^2}$  scales with a $(\tau/\tau_0)^3$, similarly to the superdiffusive regime of separation in Eq.~\eqref{eq:separation_laws}, and independently of $r_0$. This is a central result in this work, demonstrating that a superdiffusive separation law is observed at various $r_0$ and at times much smaller than $\tau_0$, despite the fact that $r_0 > \eta$.

\begin{figure*}[!ht]
	\begin{tikzpicture}
	\node[inner sep=0pt] (a) at (0,0)    
	{\includegraphics[width=0.5\textwidth]{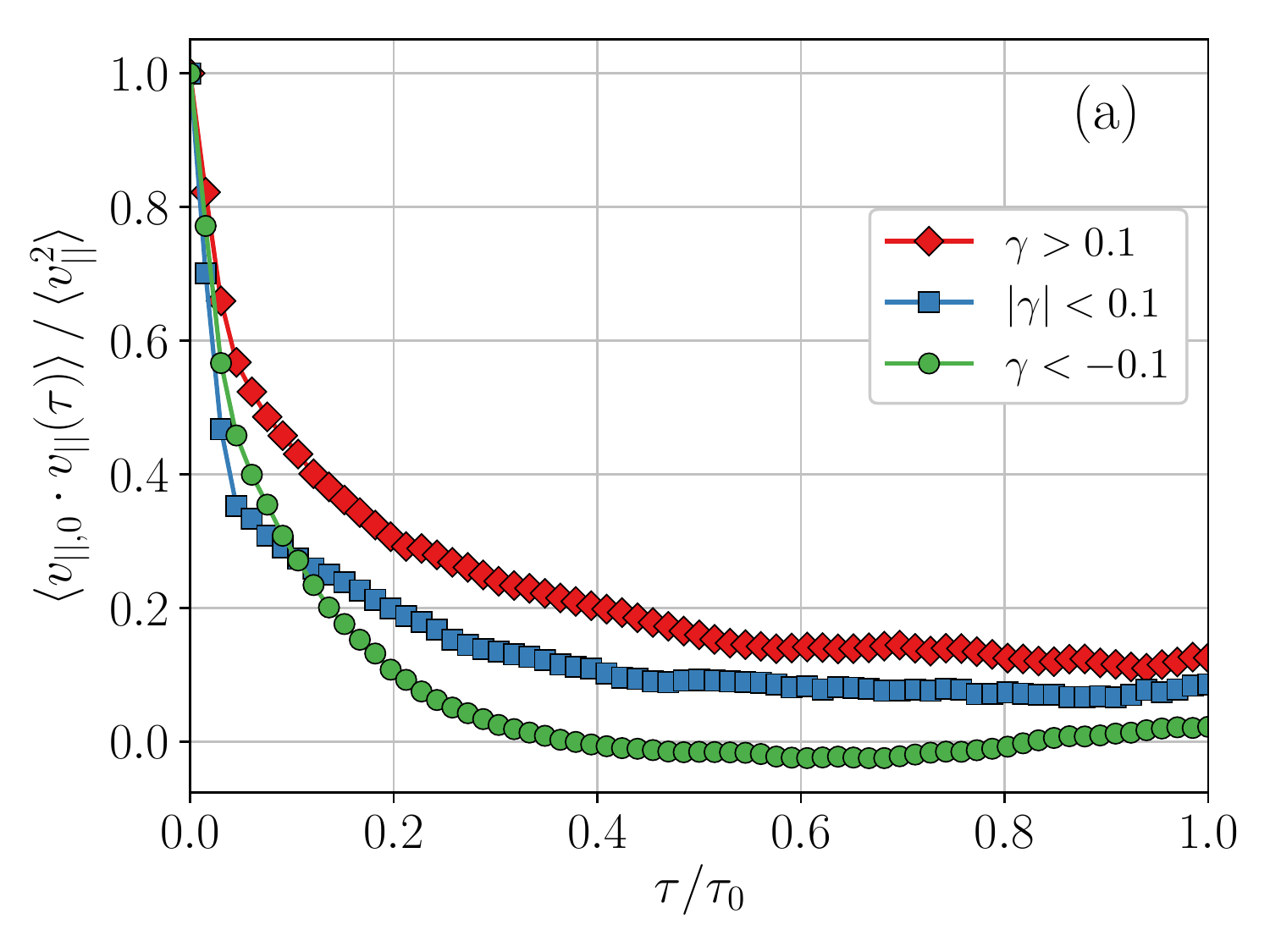}};
	
	\node[inner sep=0pt] (b) at (8.85,0) 
	{\includegraphics[width=0.5\textwidth]{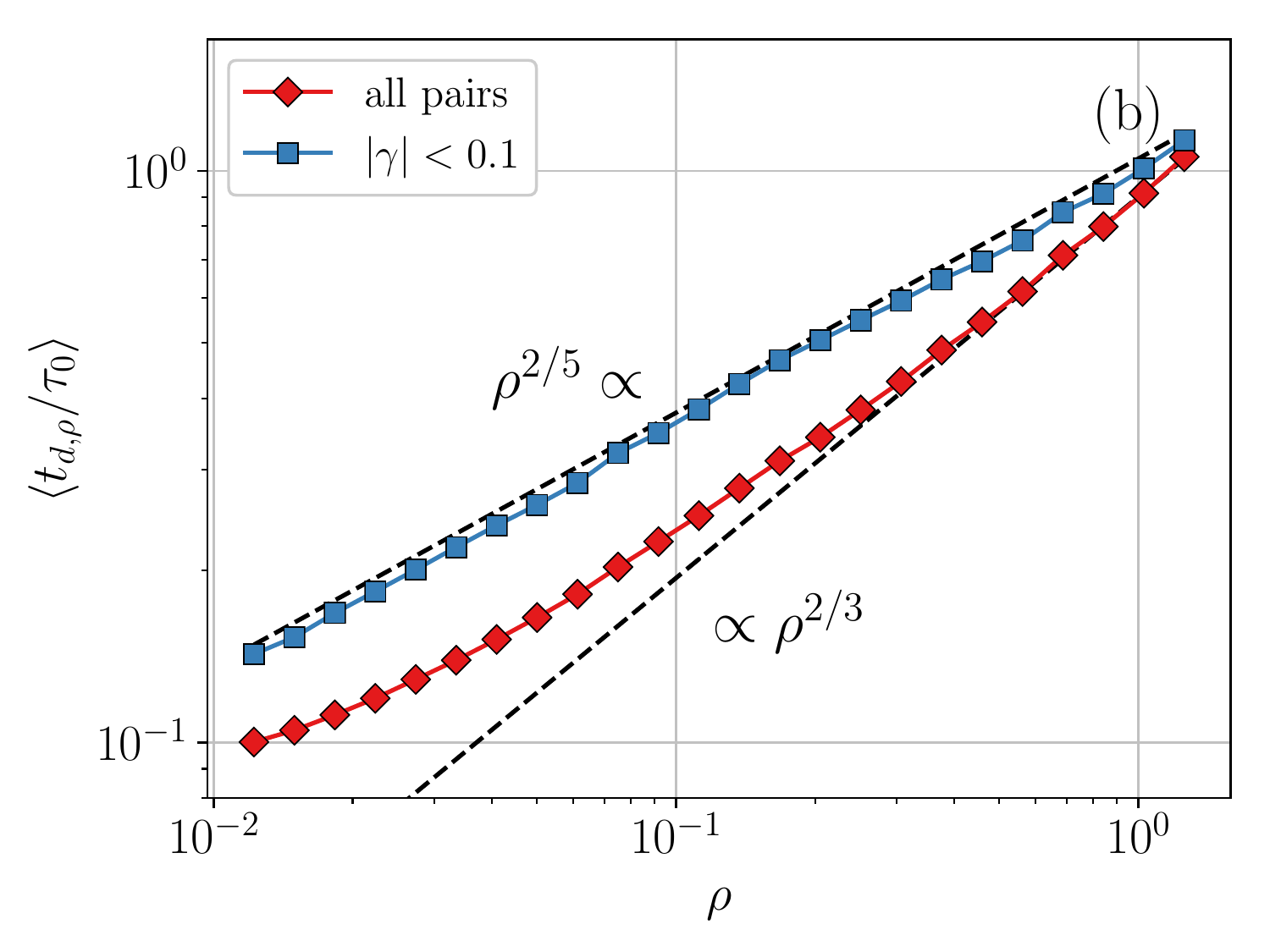}};
	\end{tikzpicture}
	
	\caption{(a) - Autocorrelation function of separation velocity, $v_{||} = dr/dt$, of pairs from three groups - large $\gamma$, negative $\gamma$ and very small $|\gamma|$. (b) - average time to reach $\Delta r = \rho r_0$ ("exit time") normalized by the eddy turnover time at scale $r_0$, plotted vs. $\rho$. Curves are shown for two groups of pairs distinguished by $|\gamma|$.  \label{fig:time_stats}}
\end{figure*}

Figure~\ref{fig:separation_curves}(b) strengthens this result, presenting $\av{\Delta r^2 /r_0^2}$ and averaged over all $r_0$, compensated by $(\tau/\tau_0)^3$ for three levels of $|\gamma|$. The graphs clearly show the effect of small/large $\tau_{v_0}$ in respect to $\tau_0$. For pairs with $|\gamma|<10$ ($\tau_0 \ll \tau_{v_0}$), a ballistic regime spans almost the entire length of the measurements. For $|\gamma| < 1$, an initial ballistic regime spans roughly up to $\tau \approx 0.1 \tau_0$, where we observe a slight slope change. Only for the ''initially passive'' pairs, for which $|\gamma| < 0.1$, we observe a long plateau that corresponds to the $\tau^3$ regime. The strict conditioning leads to a relatively small dataset of long pair trajectories, that is manifested in the loss of statistics at long times. A similar notion of extremely fast  and slow pairs was presented in Ref.~\cite{Scatamacchia2012} for $r_0 \sim \eta$. Our analysis further suggests that extreme events dominate pair dispersion from larger initial separations, since $\gamma$ is highly sensitive to the value of $v_{||,0}$.
    
Comparing the results with equation~\eqref{eq:separation_laws}, the plateau in figure~\ref{fig:separation_curves}(b) can be interpreted as a value of $g = 2.6 \pm 0.2$ for the pairs with $|\gamma|<0.1$. This value is higher than the previously reported $g \approx 0.5$ in e.g. Ref.~\cite{Julien1999,Ott2000,Boffetta2002,Biferale2005}, and can be explained in view of recent developments by Ref.~\cite{Bourgoin2015}. In Ref.~\cite{Bourgoin2015} the pair dispersion process was modeled using a cascade of ballistic events, tuned through a persistent parameter. Lower values of persistent parameter in the model exhibit higher values of the Richardson constant~\cite{Bourgoin2015}. This is analogous to our conditioning on $|\gamma|$, as will be shown in the following analysis of Lagrangian correlations.

We analyze Lagrangian autocorrelations of pair separation velocity, $v_{||}$, conditioned on $\gamma$. In figure~\ref{fig:time_stats}(a) we plot the normalized auto-correlation for the three sets of $\gamma$: negative, positive and small. The graphs show that as $\gamma$ increases, so does the correlation. Specifically, the autocorrelation for pairs that are approaching each other ($\gamma< - 0.1$) drops very rapidly to negative values and exhibits negligible correlation levels from there on. The pairs with highest $\gamma$ values, $\gamma > 0.1$, have the highest auto-correlation, that, notably, hold for the entire length of our measurement. The pairs with $|\gamma|<0.1$ show intermediate correlation values which are lower than the initially separating pairs and higher than those that were rapidly approaching each other. Therefore, Lagrangian autocorrelation results strengthen the notion that the separation of pairs with $\tau_0 \ll \tau_{v_0}$, is dominated by persistence of the rate of separation at $\tau=0$.

%

Next we turn to analyze pair dispersion in terms of fixed spatial scales, inspired by the studies of ``exit times''~\cite{Boffetta2001} or ``delay times'' analysis~\cite{Rast2011}. Because 3D-PTV measurements are limited in the range of large scales, we use "delay times", $t_{d,\rho}$, defined as the time $\tau$ it takes for a pair to cross a certain threshold scale, $\rho$, defining the next scale, $\Delta r = r_0 \rho$. It is expected that in the inertial range $\av{t_{d,\rho}}$ depends only on $r_0$, $\epsilon$ and $\rho$, or in other words: 
\begin{equation}
\av{t_{d,\rho}}  = \left( {r_0^2}/{\epsilon} \right)^{1/3}  f(\rho) =  \tau_0 \; f(\rho).
\label{eq:delay_times}
\end{equation}
\noindent The dimensionless function $f(\rho)$ should be universal in the limit of the infinite $Re$ number. Also, in analogy to the mean exit times, i.e., the analytical solution in~\cite{Boffetta2002,Biferale2005} should scale as $f(\rho) \propto \rho^{2/3}$.

For each pair in our database we have determined the delay times $t_{d,\rho}$ as a function of $\rho$ in the available range, from $0.01$ to $1.25$. Next we estimate the conditionally ensemble averaged $\av{t_{d,\rho} / \tau_0}$ for all the pairs and for the pairs with $|\gamma|<0.1$, shown in figure~\ref{fig:time_stats}(b). For all the pairs, $f(\rho)$ approaches the expected $\rho^{2/3}$ at large $\rho$ available in this experiment. However, the pairs with $|\gamma|<0.1$, scale differently, with $\av{t_{d,\rho} / \tau_0} \propto \rho^{2/5}$. 
If we assume that pair dispersion at large initial separations $r_0$ are affected by a scale-dependent dissipation rate, $\epsilon_r = \epsilon \left({r}/{L}\right)^{\mu}$ (a concept known as refined Kolmogorov similarity~\cite{Monin2007}), we arrive at the observed $\av{t_{d,\rho} / \tau_0} \propto \rho^{2/5}$. Therefore the observed scaling is, in our opinion, an evidence of a self-similar behavior of pairs with non-persistent initial rate of separation.

To conclude, in this Letter we present a generalization of the pair dispersion. \new{In previous investigations, it was shown that a superdiffusive regime of pair dispersion exists for pair separation with small initial separation. We extend this notion to large initial separations $\eta < r_0 < L$ by taking into consideration a new dimensionless parameter},\del{We show the experimental results for pair dispersion in a turbulent flow without mean shear for the pairs starting from  $\eta < r_0 < L$. We and characterize the pair dispersion in terms of the new dimensionless parameter } $\gamma = \tau_{v_0} / \tau_0$, that represents the persistence of initial conditions with respect to the eddy turnover time at $r_0$. \new{This forms an extension to the known superdiffusive separation in the case of small initial separations that was suggested by~\cite{Richardson1926}, and confirmed by~\cite{Biferale2005}.} Results from 3D-PTV show that the pairs of particles with low values of $|\gamma|$ separate superdiffusively with $\av{\Delta r^2} \propto \tau^3$, in contrast to the case of all the pairs, for which $\av{\Delta r^2}$ increases ballistically. \new{These observations are verified against high Reynolds number DNS results (see supplementary material).} Lagrangian correlation analysis of the process indicates that pairs with high $\gamma$ separate in a stronger auto-correlation as compared to the pairs with small values of $|\gamma|$. We also demonstrate for the first time that contrary to the ``delay times'' statistics of all the pairs that tends asymptotically to the expected $\rho^{2/3}$ scaling law, pairs with low $\gamma$ exhibit $\rho^{2/5}$ scaling which is consistent with the refined Kolmogorov similarity hypothesis~\cite{Monin2007}. 

\medskip
\begin{acknowledgments}
We would like to thank Micka\"{e}l Bourgoin for the useful comments on the early draft. This work was partially supported by the Pazy Research Foundation.
\end{acknowledgments}

\nocite{Li2008}
\nocite{Yu2012}

\bibliographystyle{apsrev4-1}
\bibliography{bibliography}

\end{document}